\documentclass[12pt,preprint]{aastex}
\usepackage{epstopdf}

\shorttitle{Metal-Rich M-dwarf Planet Hosts}
\shortauthors{Rojas-Ayala et al.}

\begin{document}

\title{Metal-Rich M-dwarf Planet Hosts: Metallicities with K-Band Spectra}

\author{B{\'a}rbara Rojas-Ayala\altaffilmark{1}} 
\altaffiltext{1}{Claudio Anguita Fellow}    
\and
\author{Kevin R. Covey\altaffilmark{2,3}, Philip S. Muirhead\altaffilmark{4}, and James P. Lloyd}
\affil{Astronomy Department, Cornell University,
    Ithaca, NY 14850}
\altaffiltext{2}{Hubble Fellow}
\altaffiltext{3}{Visiting Researcher, Department of Astronomy, Boston University, 725 Commonwealth Ave, Boston, MA 02215}
\altaffiltext{4}{NASA Earth and Space Science Fellow}

\email{babs@astro.cornell.edu}

\begin{abstract}

A metal-rich environment facilitates planet formation, making metal-rich stars the most favorable targets for surveys seeking to detect new exoplanets. Using this advantage to identify likely low-mass planet hosts, however, has been difficult: until now, methods to determine M-dwarf metallicities required observationally expensive data (such as parallaxes and high-resolution spectra), and were limited to a few bright cool stars. We have obtained moderate (R$\sim$2700) resolution K-band spectra of 17 M-dwarfs with metallicity estimates derived from their FGK companions. Analysis of these spectra, and inspection of theoretical synthetic spectra, reveal that an M-dwarf's metallicity can be inferred from the strength of its Na {\footnotesize I} doublet (2.206 $\mu$m \& 2.209 $\mu$m) and Ca {\footnotesize I} triplet (2.261 $\mu$m, 2.263 $\mu$m \& 2.265 $\mu$m) absorption lines. We use these features, and a temperature-sensitive water index, to construct an empirical metallicity indicator applicable for M-dwarfs with near-solar metallicities (-0.5$<$[Fe/H]$<$+0.5). This indicator has an accuracy of $\pm$0.15 dex, comparable to that of existing techniques for estimating M-dwarf metallicities, but is more observationally accessible, requiring only a moderate resolution K-band spectrum. Applying this method to 8 known M-dwarf planet hosts, we estimate metallicities ([Fe/H]) in excess of the mean metallicity of M-dwarfs in the solar neighborhood, consistent with the metallicity distribution of FGK planet hosts.

\end{abstract}


\section{Introduction}

M-dwarfs are under intense scrutiny as potential planet hosts. M-dwarfs have sufficiently low masses ($0.6M_{\sun}>M_{\star}>0.08M_{\sun}$) and small radii ($0.6R_{\sun}>R_{\star}>0.1R_{\sun}$) that exoplanets induce considerably larger reflex velocities and transit depths than an identical planet would around larger, more massive hosts. Accordingly, most of the lowest-mass planets detected to date orbit low-mass stars. M-dwarfs' relatively low luminosities also ensure that habitable planets will be in small, fast orbits, producing higher reflex radial velocity amplitudes, and a greater likelihood, depth, and frequency of transit events \citep{2008PASP..120..317N}. M-dwarfs constitute more than two thirds of the nearby stellar population \citep{2003PASP..115..763C}; identifying the most likely planet hosts in advance would greatly enhance the efficiency of M-dwarf exoplanet surveys.

Observations of FGK-dwarfs have convincingly shown that gas giant planet frequency rises steeply with host star metallicity, a result known as the planet-metallicity correlation \citep{1997MNRAS.285..403G, 2004A&A...415.1153S, 2005ApJ...622.1102F}. Determining if this correlation extends into the M-dwarf regime requires accurate metallicity estimates for ultra-cool stars. As M-dwarf spectra are dominated by chemically complex molecular features, {\em a priori} spectral synthesis does not well match the observed spectra. Most attempts to calibrate M-dwarf metallicity indicators have therefore focused on observational studies of wide binaries, where the robustly estimated [Fe/H] of the FGK-primary is assumed to describe the M-dwarf secondary as well. This assumption will hold if both binary components formed from the same well-mixed molecular cloud, and no mass transfer or dredge-up has occurred in the system. \citet[B05 hereafter]{2005A&A...442..635B} adopted this technique to develop a photometric metallicity calibration that suggested that nearby M-dwarfs, including the planet hosts GJ 876 and GJ 436, were slightly metal poor. \citet{2006ApJ...653L..65B} also estimated subsolar metallicities for GJ 436, GJ 581 and GJ 876 from spectral synthesis of optical spectra. Though there were only a small number of M-dwarf planet hosts in their samples, the results of the B05 and \citet{2006ApJ...653L..65B} studies were unexpected, since they imply that planet formation around M-dwarfs does not depend on the star's metal content. This finding is difficult to reconcile with the planet-metallicity correlation for FGK-dwarfs, and runs counter to the predictions of the commonly accepted core accretion model \citep{1974Icar...22..416P}.

\citet[JA09 hereafter]{2009ApJ...699..933J} derived a new photometric metallicity calibration using 6 M-dwarfs with wide, metal-rich FGK companions not present in the B05 study. JA09 found that the previous photometric calibration systematically underestimated the metallicities of these metal-rich stars. JA09 attributed this discrepancy to the lack of metal-rich calibrators in the B05 sample and/or the use of visual magnitudes with poorly characterized uncertainties. JA09 also estimated the metallicities of 7 M-dwarfs with planetary-mass companions, concluding that M-dwarf planet hosts are preferentially metal-rich, just like FGK planet hosts. JA09 suggested that the lack of Jovian planets around M-dwarfs is most likely due to the influence of stellar mass, not stellar metallicity, on the planet formation process, consistent with the work of \citet{2004ApJ...612L..73L} on giant planet formation around red dwarfs. 

\citet[SG10 hereafter]{2010arXiv1006.2850S} recently revised the B05 and JA09 photometric calibrations, finding that they systematically underestimate or overestimate metallicity at the extremes of their range. Their results hint that low-mass planets may be more likely to be found around metal-rich M-dwarfs, which disagrees with the almost equal probability of finding Neptune-mass planets around a metal-rich or a metal-poor FGK-star \citep{2008A&A...487..373S}.

Existing methods for estimating M-dwarf metallicities rely heavily on observations at visual wavelengths, where metallic absorption lines and molecules strongly influence a star's spectrum. Mid- to late- M-dwarfs are typically too faint for high-quality observations at visual wavelengths, which has limited most metallicity analyses to early-M-dwarfs and a few nearby stars \citep[e.g. ][]{2005MNRAS.356..963W}. Lacking a satisfactory way to estimate metallicities across the entire M-dwarf Sequence has limited our understanding of planet formation around the lowest mass stars.

In this Letter, we present a technique for estimating the metallicity of field M-dwarfs from the strength of prominent K-band absorption features. We calibrated our technique using spectra of M-dwarfs with existing [Fe/H] estimates, adopted as a proxy for their overall metallicities, obtained from the analysis of a wide FGK-companion. Our technique does not depend on accurate parallaxes and V magnitudes, or high-resolution, high-signal-to-noise spectra, as required by previous techniques. Instead, our technique requires simply a K-band spectrum at moderate resolution, which can be efficiently obtained for many M-dwarfs with current near-infrared spectrographs. Our method will enable metallicity estimates to be generated for stars which are too cool or too distant for other methods, identify the most advantageous targets for planet-searches around cool stars, and thus improve our understanding of planet formation around low-mass stars. In \S 2, we describe our observations and data reduction. We present our analysis and results in \S 3, and summarize and discuss our conclusions in \S 4.

\section{Observations and Data Reduction}

Near-infrared spectra of the stars in this work were obtained with the TripleSpec spectrograph on the Palomar Hale Telescope \citep{2008SPIE.7014E..30H} during 7 observing runs between 2007 and 2009. TripleSpec at Palomar has no moving parts and simultaneously acquires 5 cross-dispersed orders covering 1.0-2.4 $\mu$m at a resolution of $\lambda/\Delta\lambda$$\approx$2700.

The spectra were reduced with an IDL-based data reduction pipeline developed by P. Muirhead\footnote{http://www.astro.cornell.edu/$\sim$muirhead/\#Downloads}. The data were sky-subtracted using a sky-frame made by median combining 5 exposures with the science target placed at different positions along the slit. Each sky-subtracted exposure was then divided by a normalized flat-field, wavelength calibrated and optimally extracted \citep{1986PASP...98..609H}. The spectra were flux-calibrated and telluric corrected using observations of an A0V star at a similar airmass with the IDL-based code $xtellcor$\_$general$ by \citet{ 2003PASP..115..389V}. 

Our sample consists of:

\begin{itemize}
\item Seventeen M-dwarfs with wide ($>$ 5$\arcsec$ separation), common-proper-motion solar-type companions to serve as metallicity calibrators. The FGK-primaries have spectroscopic metallicity measurements by \citet[SPOCS Catalogue]{2005ApJS..159..141V}. The binary systems were selected from the \citet{1991STIA...9233932G} catalogue of nearby stars, the \citet{1994RMxAA..28...43P} catalogue of nearby wide binary and multiple systems, and the list of new HIPPARCOS binaries by \citet{2004ApJS..150..455G}. 

\item Twenty-nine nearby M-dwarfs analyzed by JA09, to compare our spectroscopic results with their photometric [Fe/H].

\item Eight M-dwarf planet hosts observable from Palomar Mountain: GJ 876 \citep{1998ApJ...505L.147M,1998A&A...338L..67D}, GJ 436 \citep{2004ApJ...617..580B,2004A&A...426L..19S}, GJ 581 \citep{2005A&A...443L..15B}, GJ 849 \citep{2006PASP..118.1685B}, GJ 1214 \citep{2009Natur.462..891C}, GJ 649 \citep{2010PASP..122..149J}, HIP 57050 \citep{2010ApJ...715..271H}, and HIP 79431 \citep{2010PASP..122..156A}.  

\item Twenty-nine M-dwarf spectroscopic standards defined by \citet[KHM system]{1991ApJS...77..417K} or with KHM spectral types from \citet{1994AJ....108.1437H}

\end{itemize}

\begin{figure*}
\epsscale{0.9}
\plotone{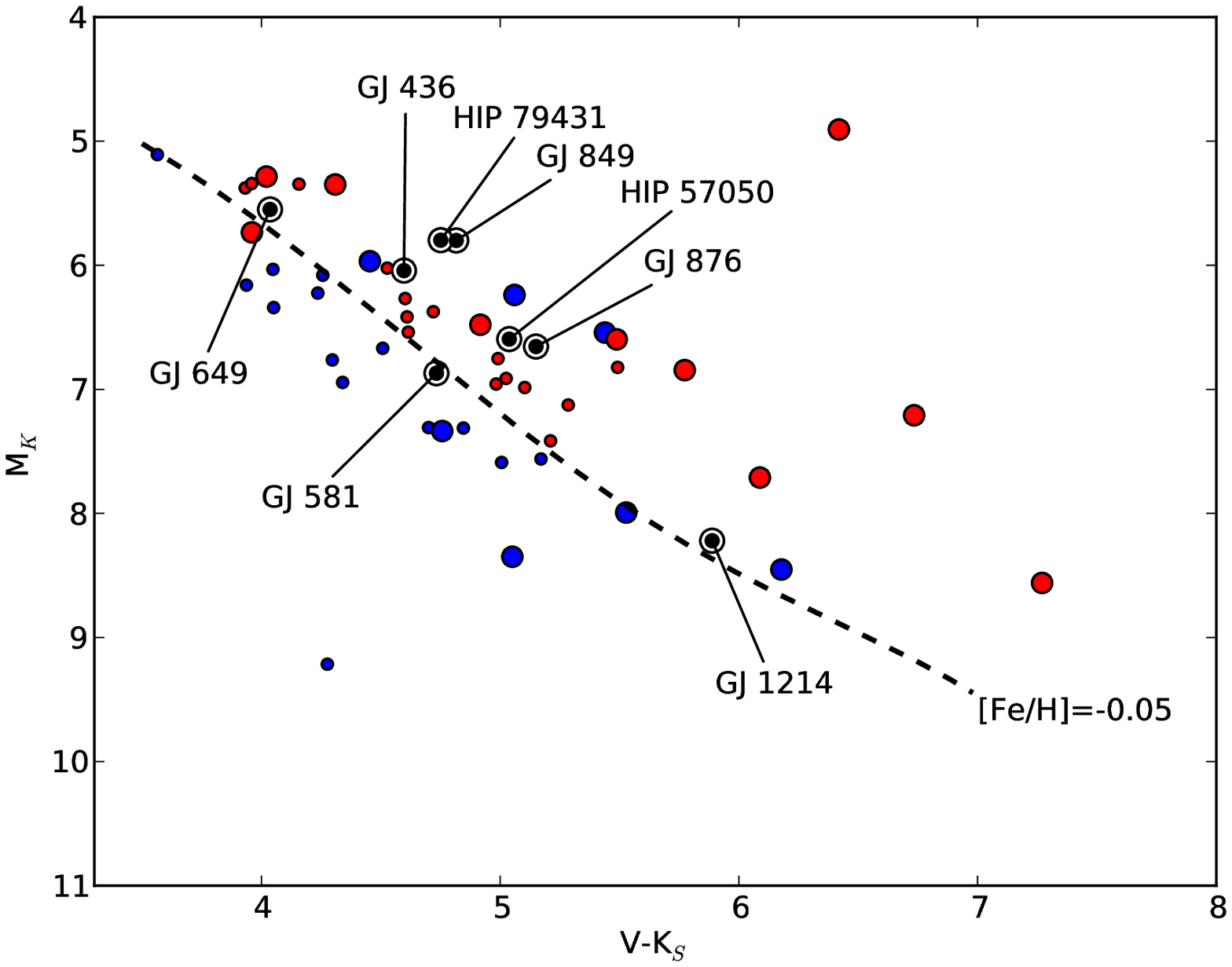}
\caption{ The $\{$V - K$_S$, M$_K$$\}$ plane for our M-dwarf sample. The big red and blue dots are M-dwarfs in our metallicity calibration sample with [Fe/H]$>$-0.05 and [Fe/H]$<$-0.05, respectively. The black dotted line is a fifth-order polynomial fit to the mean Main Sequence calculated by JA09 ([Fe/H]=-0.05). The small red and blue dots represent M-dwarfs with [Fe/H] $\geq$-0.05 and [Fe/H] $<$-0.05 respectively, according to the photometric calibration by JA09. The big black dots represent the M-dwarf planet hosts. The lack of  precise parallaxes for most of the metallicity calibrators can explain the position of three metal-poor M-dwarf secondaries above the grey-dotted line, where metal-rich M-dwarfs are supposed to be according to JA09.}
\label{fig1}
\end{figure*}

Figure \ref{fig1} shows the $\{$V-K$_S$, M$_K$$\}$ plane for our sample. The JA09 calibration is only valid for stars with: 1) accurate Johnson V magnitudes and parallaxes from the Hipparcos Catalogue \citep{2008yCat.1311....0V} , the Yale Parallax Catalogue \citep{2001yCat.1238....0V} or the TASS Mark IV Survey Catalogue \citep{2007yCat.2271....0D}, and 2) V-K$_S$ colors between [3.9,6.6]. JA09 assume that the mean M$_K$ versus V-K$_S$ relation in the solar neighborhood is equivalent to an [Fe/H]=-0.05 metallicity contour, with stars brighter than this being metal-rich, and fainter stars being metal-poor. Most of the mid-type and all of the late-type M-dwarfs fail both of these conditions: they lack accurate parallaxes and/or Johnson V magnitudes, so were not included in the JA09 calibration.

\section{Analysis and Results}

\begin{figure}
\epsscale{0.7}
\plotone{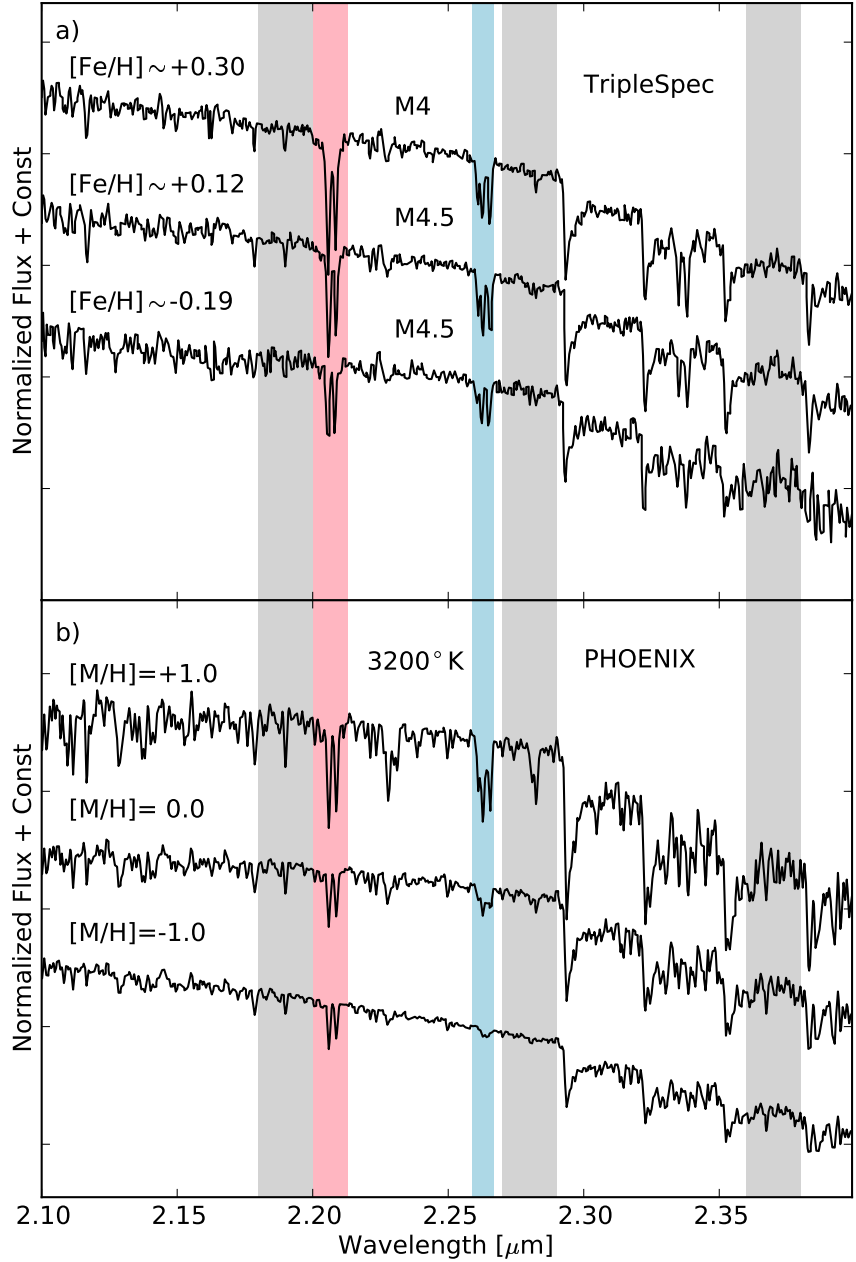}
\caption{ (a) K-band TripleSpec spectra of GJ 324 B (top), HIP 57050 (middle), and  GJ 783.2 B (bottom). (b) Updated PHOENIX models with T$_{eff}$=3200 K, log g=4.5, and varying overall metallicities. The regions used to calculate the EWs of the Na {\footnotesize I} and the Ca {\footnotesize I} features, and the H$_2$O-K index defined by \citet{2010arXiv1007.2192C} are shown in red, blue and grey, respectively. The updated PHOENIX models are consistent with the overall shape of the M-dwarf spectra. However, there are discrepancies in the strength of absorption features between the M-dwarf spectra and the models of similar metallicity.}
\label{fig2}
\end{figure}

We have empirically calibrated several NIR spectral features to serve as effective M-dwarf metallicity indicators. Figure \ref{fig2} shows the K-band spectra of three M4-M4.5 dwarfs in our sample, along with updated PHOENIX models \footnote{T. Barman, private communication} \citep{1999ApJ...512..377H} with effective temperatures equivalent to an $\sim$M3.5-4 star and varying metallicities. The overall shape of the K-band spectrum is well represented by the updated PHOENIX models, and the Na {\footnotesize I} doublet (2.206 $\mu$m \& 2.209 $\mu$m) and the Ca {\footnotesize I} triplet (2.261 $\mu$m, 2.263 $\mu$m \& 2.265 $\mu$m) are seen to strengthen considerably with metallicity. There is a discrepancy, however, between the strengths of these features in our spectra and in the models of comparable metallicity.

We measured the equivalent width (EW) of the Na {\footnotesize I} and the Ca {\footnotesize I} features to differentiate between metal-rich and metal-poor M-dwarfs. Each spectrum was normalized and the EWs were calculated by computing the ratio of the area of a feature to a pseudocontinuum across that feature, computed from featureless regions on either side of it, using the the IDL-based function \verb*#measure_ew# by N. Konidaris and J. Harker. Uncertainties in the EWs were obtained from the errors in the estimated pseudocontinuum and the errors in the measured intensities of each feature, as described in \citet{1992ApJS...83..147S}.

\begin{figure}
\epsscale{0.7}
\plotone{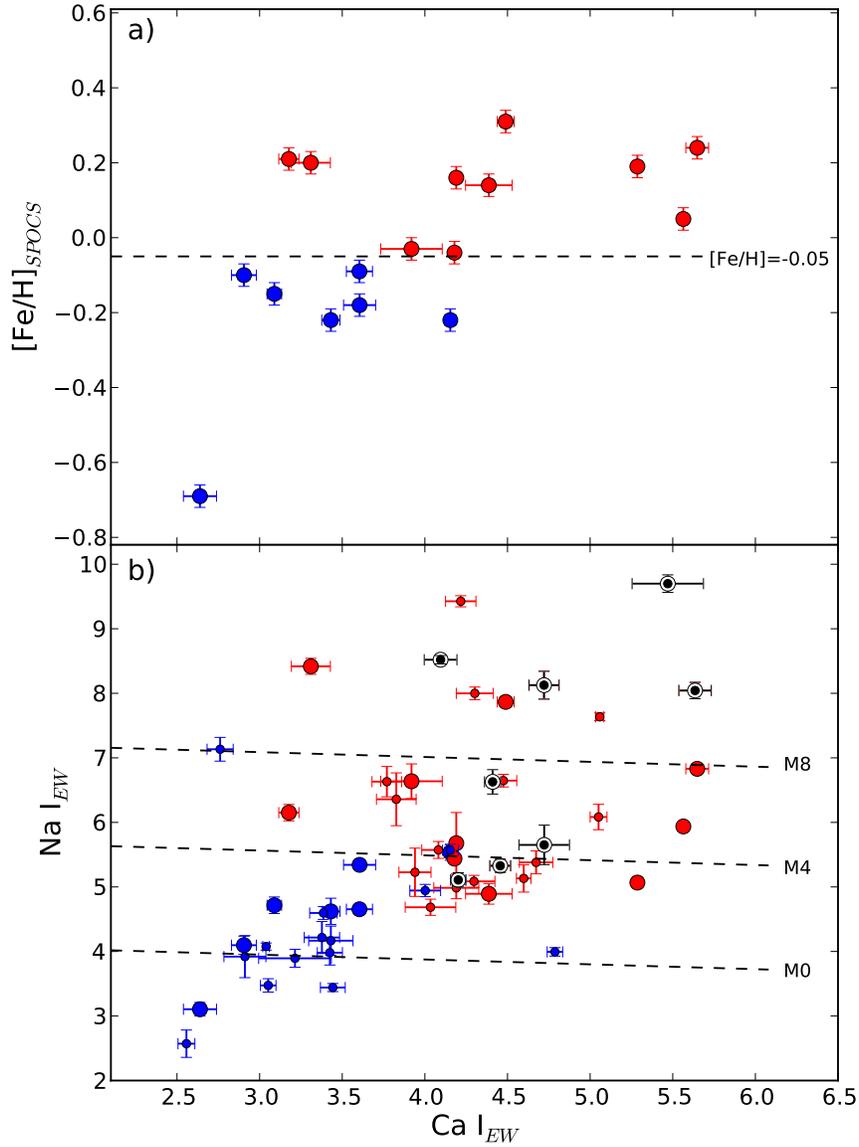}
\caption{(a) SPOCS [Fe/H] versus the EWs of the Ca{\footnotesize I} triplet for the M-dwarfs in the metallicity calibration sample. The nomenclature for the stars is the same as in Figure \ref{fig1}. The dashed line is the mean metallicity of the solar neighborhood by JA09 ([Fe/H]=-0.05). (b) The EW of Na {\footnotesize I} versus the EW of Ca {\footnotesize I} features for our M-dwarf sample. The dashed lines are the [Fe/H]=-0.05 contours for the spectral types M0, M4 and M8, calculated from Equation (2). These alkali features correlate with [Fe/H], but the dispersion seen is partially due to differences in the stars' temperature: an M8-dwarf has larger EW values than an M0-dwarf of the same metallicity.}.  
\label{fig3}
\end{figure}

Figure \ref{fig3}a compares the SPOCS [Fe/H] for our M-dwarf metallicity calibrators to the EWs of their Ca {\footnotesize I} triplets; Figure \ref{fig3}b compares the EWs of the Na {\footnotesize I} and the Ca {\footnotesize I} features for our M-dwarf sample. The correspondence between the Na {\footnotesize I} and Ca {\footnotesize I} features and the metallicity of M-dwarfs is evident: metal-poor M-dwarfs have lower EWs than metal-rich M-dwarfs. The strength of these features also depend on temperature and surface gravity, however, as do most atomic lines in the NIR spectra of M-dwarfs \citep{1995AJ....110.2415A, 2004ApJS..151..387I}.  

To explicitly account for the influence of stellar temperature on the strengths of Na and Ca lines, we include an independent, temperature sensitive spectral index in our analysis. K-band H$_2$O absorption bands been used extensively to diagnose M-dwarf spectral types \citep[e.g., ][]{2003ApJ...596..561M}; we diagnose the effective temperatures of stars in our sample using the H$_2$O-K index defined by \citet{2010arXiv1007.2192C}:

\begin{eqnarray*}
\mathrm{H_{2}O\!\!-\!\!K}=\frac{\langle \mathcal{F}(2.18-2.20) \rangle / \langle \mathcal{F}(2.27-2.29) \rangle}{\langle \mathcal{F}(2.27-2.29) \rangle / \langle \mathcal{F}(2.36-2.38) \rangle}
\end{eqnarray*}

\noindent where $\langle \mathcal{F}(a-b) \rangle$ denotes the mean flux level in the wavelength range defined by $a$ and $b$, in microns. The linear relation between H$_2$O-K index and the KHM spectral types of spectral standards in our sample is: 

\begin{eqnarray}
\mathrm{Sp.Type} &=& 42.343 -42.920 (\mathrm{H_{2}O\!\!-\!\!K}) \\
\sigma\mathrm{(Sp.Type)} &=&0.94 \nonumber 
\end{eqnarray}
 
 \noindent We report the spectral types estimated from Equation (1) for the metallicity calibration M-dwarfs and planet hosts in Table 1, rounded to the nearest half subtype.

A star's surface gravity also influences its spectrum, but those are typically subtle effects. All of our targets are dwarf stars, which reduces any potential gravity ambiguity in our analysis. \citet{2009ApJ...701..764F} reported empirical values of log g$\sim$4.9 for five M-dwarfs in eclipsing binary systems, while values of log g$\sim$4.6-4.9 have been inferred from mass and radius values for M-dwarfs in \citet{2009A&A...505..205D}. 

To identify the best fit relationship between [Fe/H], the EWs of Na {\footnotesize I} and Ca {\footnotesize I} features in $\AA$, and the H$_2$O-K index, we performed a linear regression on the parameters measured for our M-dwarfs with SPOCS metallicities. We found a best-fit linear equation of:

\begin{eqnarray}
\mathrm{[Fe/H]} &=&~0.142(\mathrm{Na~{\footnotesize I}_{EW}})~+~0.011(\mathrm{Ca~{\footnotesize I}_{EW}}) \nonumber \\
&&~+~2.541(\mathrm{H_{2}O\!\!-\!\!K})~-~3.132 \\
\sigma\mathrm{([Fe/H])} &=&0.15 \nonumber 
\end{eqnarray}

\noindent with a residual mean square $RMS$([Fe/H])=0.02 and an adjusted squared multiple correlation coefficient $R$$^2_a$([Fe/H])=0.63, which implies, by comparison of the same criteria functions calculated in SG10, that our model provides a better fit than the models of B05 ($RMS$=0.05,  $R$$^2_a$$<$0.05), JA09 ($RMS$=0.04,  $R$$^2_a$$=$0.059) and SG10 ($RMS$=0.02,  $R$$^2_a$$=$0.49).  We estimate systematic errors, such as the zero-point value and the inclusion of a non-physically bound binary as a metallicity calibrator, to be of the order of $\sim$0.06 (SPOCS data) and $\sim$0.03, respectively. The combined systematic error is of the same order of the structure in the [Fe/H] residuals of the calibration sample. We note that the predicted [Fe/H] for some of the planet hosts lie above the [Fe/H] boundaries of the calibrators. The values of the EWs of Na {\footnotesize I} and Ca {\footnotesize I}, the H$_2$O-K index and [Fe/H] values for the metallicity calibration M-dwarfs and planet hosts, are given in Table 1.

The best fit linear combination of the EWs of the Ca {\footnotesize I} and the Na {\footnotesize I} features are plotted in Figure \ref{fig4}a versus the H$_2$O-K index of our M-dwarf sample. There is a clear distinction between the M-dwarfs with metal-rich and metal-poor FGK-dwarf companions in Figure \ref{fig4}a; the SPOCS [Fe/H] values of our M-dwarf metallicity calibrators are also plotted in Figure \ref{fig4}b against the value of their H$_2$O-K index. The M-dwarfs with photometric metallicity estimates calculated by JA09 are also bisected between metal-rich and metal-poor using these K-band features. Moreover, we found that all of the M-dwarf planet hosts have metallicities much higher than the mean metallicity of M-dwarfs in the solar neighborhood ([Fe/H]$_{SN}$$\sim$-0.17), defined by SG10. The Jovian planet hosts are co-located with the M-dwarf companions of metal-rich FGK stars in the upper portion of Fig. \ref{fig4}a, while the Neptune planet hosts are located just above the M-dwarf companions of metal-poor FGK stars.

\section{Discussion}

\begin{figure}
\epsscale{1}
\plotone{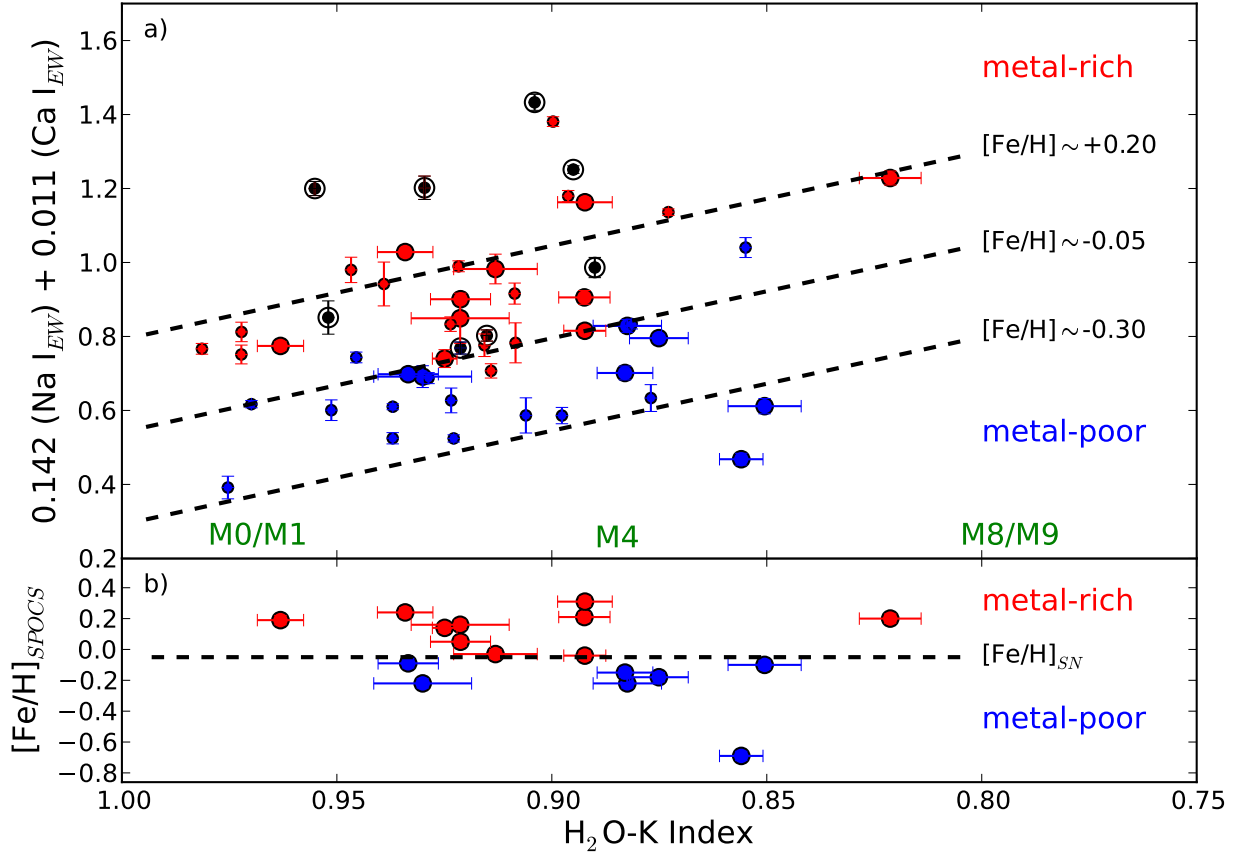}
\caption{A linear combination of the EWs of the Ca {\footnotesize I} and Na {\footnotesize I} features versus the H$_2$O-K index for our M-dwarf sample. The nomenclature for the stars is the same as in Figure \ref{fig1}. The [Fe/H] values for the metallicity calibrators are also plotted versus the H$_2$O-K index, to emphasize the index's insensitivity to metallicity. The dashed lines in the top panel are iso-metallicity contours for [Fe/H] values of -0.30, -0.05 and +0.20, calculated from Equation (2). According to our determination, the Jovian M-dwarf planet hosts have higher metallicities than M-dwarfs with Neptune- or Earth-size planets, which is in agreement with the metallicity distribution of FGK-dwarfs with planets.}
\label{fig4}
\end{figure}

We have found that the EWs of the Ca {\footnotesize I} triplet, the Na {\footnotesize I} doublet, and water absorption in the K-band differentiate metal-rich and metal-poor M-dwarfs, as shown in Figure \ref{fig4}. We estimate [Fe/H] values higher than -0.05 dex for eight M-dwarf planet hosts, with the Jovian planet hosts being more metal-rich than their Neptune host analogs (Table 1). Our results, along with the conclusions of JA09 and SG10 that M-dwarf planet hosts appear to be systematically metal-rich, are consistent with the metallicity distribution of FGK-dwarfs with planets. 

We also estimated spectral types for the M-dwarf planet hosts and metallicity calibrators in our sample, using the H$_2$O-K index defined by \citet{2010arXiv1007.2192C}.  Our spectral type estimates agree well with the KHM spectral types \citep{1991ApJS...77..417K}, modulo the $\sim$1 subtype errors (Table 1). We do find discrepancies, however, with the spectral types of some of the most metal-rich and metal-poor stars in our sample. The metal-rich planet hosts GJ 849 and GJ 876 have later spectral types in the KHM system than are indicated by the H$_2$O-K index. A possible explanation is that the KHM system relies strongly on the double-metal TiO and VO molecular bands in the 6300-9000 $\AA$ wavelength range. These molecular bands are known to be metallicity dependent \citep[e.g. the subdwarf classification by][]{1997AJ....113..806G}: they grow stronger with increasing metallicity as well as decreasing temperature. While the KHM system includes the overall shape of the stellar spectrum, this does not change considerably for early M-dwarfs, such that metal-rich early-type M-dwarfs with strong molecular features could be easily mistaken for less-metal rich later-type M-dwarfs. This reasoning can also be applied to GJ 611 B, the most metal-poor star in our sample:  we derive an M6 spectral type from its H$_2$O-K index, while its KHM spectral type is M4, consistent with its weak TiO and VO bands being interpreted as a temperature effect rather than a sign of its low metallicity.

Measuring M-dwarf metallicities will also improve our ability to determine key physical parameters for both stars and planets. The masses and radii of M-dwarf planet hosts are typically derived from theoretical models, and/or empirical relationships between these parameters and the star's temperature and luminosity, all of which have a strong dependence on metallicity \citep{1998A&A...337..403B, 1998A&A...338L..67D,2009A&A...505..205D}. Since minimum planetary masses inferred from RV measurements are tied to the mass of the host star, and planetary radii derived from transit observations depend directly on the adopted stellar radius, errors in stellar parameters lead directly to errors in inferred planetary properties.  \citet{1998A&A...338L..67D} report a stellar mass of 0.3 M$_\odot$ for GJ 876, but with an uncertainty of $\sim$30$\%$ due primarily to the uncertainty in the host's metallicity, which affected the initial estimate of the masses of its planetary companions. Recent measured M-dwarf radii and masses have typical errors less than 10$\%$ \citep{2009ApJ...701..764F}, but the errors may increase somewhat for stars with significantly non-solar metallicities. With poor estimates of the mass and radius of exoplanets, it is harder to constrain their true composition, atmospheres and interiors.

Despite the fact that K-band alkali features are temperature/gravity dependent \citep{1995AJ....110.2415A, 2004ApJS..151..387I}, and prone to non-local thermodynamic equilibrium effects, our empirical analysis demonstrates that the dispersion in EW of the Ca {\footnotesize I} and Na {\footnotesize I} features can diagnose the metal abundance of M-dwarfs with roughly the same temperature (H$_2$O-K index) and gravity\footnote{Future studies that may suffer contamination by giants should use other gravity sensitive features (e.g. the 2.296 $\mu$m CO band-head) to distinguish between dwarfs and giants prior to interpreting Na {\footnotesize I} and Ca {\footnotesize I} feature strengths as metallicity indicators.}.  Our method has an advantage with respect to previous photometric and optical spectroscopic calibrations, since it does not depend on parallaxes or V magnitudes, allowing us to cover a larger sample of cooler and distant M-dwarfs. Our results enable the identification of the most advantageous M-dwarfs for planetary search surveys around cool stars, such as the Triplespec Exoplanet Discovery Instrument \citep[TEDI]{2007SPIE.6693E..26E}, MEarth \citep{2009IAUS..253...37I}, the Precision Radial Velocity Spectrograph \citep[PVRS]{2008SPIE.7014E..31J}, and the CRIRES Survey by \citet{2010ApJ...713..410B}, and will thus help unravel the formation mechanisms of planets around M-dwarfs.

\acknowledgments

We thank the staff and telescope operators of Palomar Observatory for their support. We also thank Kevin Apps, John Johnson, John Bochansky, Andrew West and the anonymous referee for their useful comments. This publication makes use of the SIMBAD database, operated at CDS, Strasbourg, France and the data products from the 2MASS, which is a joint project of the University of Massachusetts and the IPAC/CALTECH funded by the NASA and the NSF. This work was supported by CONICYT and the AURA, Inc. in representation of the NSF under cooperative agreement NSF AST-9613615. Support for this work was provided by NASA through Hubble Fellowship grant HST-HF-51253.01-A awarded by the STScI, which is operated by the AURA, Inc., for NASA, under contract NAS 5-26555. This work was supported by NASA Headquarters under the NESSF Program - Grant NNX07AP56H. This material is based in part on work supported by the NSF under grant AST-0905932.

\clearpage

\begin{deluxetable}{lcccccccccccc}
\tabletypesize{\scriptsize}
\rotate
\tablecolumns{13}
\tablewidth{0pt}
\tablecaption{M-Dwarf Planet Hosts and Metallicity Calibration Sample}
\tablehead{
\colhead{}    &  \multicolumn{2}{c}{EW [$\AA$] } &   \colhead{Index}   &
\multicolumn{2}{c}{This Work} & \colhead{KHM}  & \colhead{SPOCS} & \colhead{B05} & \colhead{JA09} & \colhead{SG10} & \colhead{M sin(i)} & \colhead{} \\
\cline{2-3} \cline{5-6}
\colhead{Name} & \colhead{Na I}   & \colhead{Ca I}    & \colhead{H2O} &
\colhead{Sp. Type}    & \colhead{[Fe/H]} & \colhead{Sp. Type} & \colhead{[Fe/H]} & \colhead{[Fe/H]} & \colhead{[Fe/H]} & \colhead{[Fe/H]}  &\colhead{M$_J$}   & \colhead{Planet Notes} }   
\startdata
HIP 79431 		&9.699 			&5.470 		&0.904  		&M3.5 		&+0.60 		&M3.0 		&\nodata		&+0.16	&+0.52	&+0.35	&2.1  	& Single Jupiter\\
GJ 849 			&8.043 			&5.635 		&0.955 		&M1.5 		&+0.49 		&M3.5		&\nodata		&+0.14	&+0.58	&+0.41	&0.82	&Single Jupiter\\
GJ 876			&8.126 			&4.721 		&0.930 		&M2.5 		&+0.43		&M4.0	 	&\nodata		&+0.03	&+0.37	&+0.23	&2.64	&2 Jupiters + Neptune + Super Earth\\	
GJ 1214 			&8.520 			&4.095 		&0.895 		&M4.0 		&+0.39		&M4.5		&\nodata	 	&\nodata	&+0.03	&+0.28	&0.0179 	&Single Super Earth\\
GJ 649 			&5.651 			&4.722 		&0.952 		&M1.5 		&+0.14		&M1.0 		&\nodata		&-0.18	&+0.04	&-0.03	&0.328 	&Single Neptune\\
HIP 57050 		&6.628 			&4.410 		&0.890 		&M4.5 		&+0.12		&M4.0 		&\nodata 		&-0.02	&+0.32	&\nodata	&0.298 	&Single Neptune\\
GJ 436 			&5.328 			&4.456 		&0.915 		&M3.0 		&-0.00		&M2.5		&\nodata	 	&-0.03	&+0.25	&+0.10	&0.072  	&Single Neptune\\
GJ 581 			&5.108 			&4.202 		&0.921 		&M3.0 		&-0.02		&M3.0 		&\nodata  		&-0.25	&-0.10	&-0.22	&0.0492  	&Neptune + 3 Super Earths\\
\\
GJ 212			&5.067			&5.286		&0.963		&M1.0		&+0.09		&M0.5		&+0.19		&-0.05	&\nodata	&+0.09 		&\nodata &\nodata \\
HD 46375 B		&6.830 			&5.648		&0.934		&M2.0		&+0.27		&M1.0		&+0.24 		&-0.33	&+0.22	&+0.12		&\nodata &\nodata \\
GJ 797 B			&4.654			&3.604		&0.933		&M2.5		&-0.06		&M2.5	        	&-0.09 		&-0.09	&\nodata	&+0.11		&\nodata&\nodata \\
GJ 872 B			&4.621			&3.431		&0.930		&M2.5		&-0.08		&M1.0		&-0.22 		&+0.12	&\nodata	&\nodata		&\nodata &\nodata \\
GJ 250 B			&4.893			&4.387		&0.925		&M2.5		&-0.04		&M2.0		&+0.14 		&+0.12	&\nodata	&-0.07		&\nodata &\nodata \\
GJ 768.1 B		&5.677			&4.190		&0.921		&M3.0		&+0.06		&M3.5		&+0.16 		&+0.16	&\nodata	&+0.22		&\nodata &\nodata \\
NLTT 14186		&5.937			&5.564		&0.921		&M3.0		&+0.11		&M		        	&+0.05 		&\nodata	&\nodata	&\nodata		&\nodata &\nodata \\
HD 222582 B		&6.637			&3.919		&0.913		&M3.0		&+0.17		&M3.5		&-0.03 		&-0.04	&\nodata	&\nodata		&\nodata &\nodata \\
GJ 777 B			&6.151			&3.177		&0.892		&M4.0		&+0.04		&M4.5	        	&+0.21 		&\nodata	&+0.19	&\nodata		&\nodata &\nodata \\
GJ 231.1 B(C)		&5.439			&4.179		&0.892		&M4.0		&-0.05		&M3.5		&-0.04 		&+0.16	&\nodata	&+0.12		&\nodata &\nodata \\
GJ 324 B			&7.867			&4.489		&0.892		&M4.0		&+0.30		&M4.0	        	&+0.31 		&\nodata	&+0.33	&+0.26		&\nodata &\nodata \\
GJ 783.2 B		&4.716			&3.089		&0.883		&M4.5		&-0.19		&M4.0		&-0.15 		&-0.35	&\nodata	&-0.19		&\nodata &\nodata \\
GJ 3348 B		&5.534			&4.154		&0.882		&M4.5		&-0.06		&M4.0	        	&-0.22 		&+0.16	&\nodata	&\nodata		&\nodata &\nodata \\
GJ 544 B			&5.342			&3.605		&0.875		&M5.0		&-0.11		&M6.0		&-0.18 		&\nodata	&\nodata	&\nodata		&\nodata &\nodata \\
GJ 611 B			&3.105			&2.639		&0.856		&M6.0		&-0.49		&M4.0	        	&-0.69 		&\nodata	&\nodata	&\nodata		&\nodata &\nodata \\
NLTT 15867		&4.097			&2.905		&0.851		&M6.0		&-0.36		&\nodata		&-0.10 		&\nodata	&\nodata	&\nodata		&\nodata &\nodata \\
GJ 376 B			&8.418			&3.310		&0.821		&M7.5		&+0.18		&M5.0		&+0.20 		&\nodata	&\nodata	&\nodata		&\nodata &\nodata \\
\enddata
\end{deluxetable}

\end{document}